\documentclass[twocolumn,showpacs,superscriptaddress,floatfix]{revtex4}
\usepackage[intlimits]{amsmath}
\usepackage{graphicx}
\usepackage{color}
\usepackage{amstext}

\renewcommand{\Re}{\mathrm{Re}}

\begin{document}

\title{Time-irreversibility of the statistics of a single
  particle in a compressible turbulence}

\author{Tobias \surname{Grafke}}
\affiliation{Department of Physics of Complex Systems, Weizmann
  Institute of Science, Rehovot 76100, Israel}
\author{Anna \surname{Frishman}}
\affiliation{Department of Physics of Complex Systems, Weizmann
  Institute of Science, Rehovot 76100, Israel}
\author{Gregory \surname{Falkovich}}
\affiliation{Department of Physics of Complex Systems, Weizmann
  Institute of Science, Rehovot 76100, Israel}

\date{\today}

\begin{abstract}
  We investigate time-irreversibility from the point of view of a
  single particle in Burgers turbulence.  Inspired by the recent work
  for incompressible flows [Xu et al., PNAS 111.21 (2014) 7558], we
  analyze the evolution of the kinetic energy for fluid markers and
  use the fluctuations of the instantaneous power as a measure of
  time-irreversibility.  For short times, starting from a uniform
  distribution of markers, we find the scaling
  $\left<[E(t)-E(0)]^n\right>\propto t$ and $\left<p^n\right> \propto
  \Re^{n-1}$ for the power as a function of the Reynolds number. Both
  observations can be explained using the ``flight-crash'' model,
  suggested by Xu et al.  Furthermore, we use a simple model for
  shocks which reproduces the moments of the energy difference
  including the pre-factor for $\left<E(t)-E(0)\right>$.  To complete
  the single particle picture for Burgers we compute the moments of
  the Lagrangian velocity difference and show that they are
  bi-fractal. This arises in a similar manner to the bi-fractality of
  Eulerian velocity differences.  In the above setting,
  time-irreversibility is directly manifest as particles eventually
  end up in shocks. We additionally investigate time-irreversibility
  in the long-time limit when all particles are located inside shocks
  and the Lagrangian velocity statistics are stationary. We find the
  same scalings for the power and energy differences as at short times
  and argue that this is also a consequence of rare ``flight-crash''
  events related to shock collisions.
\end{abstract}



\maketitle

\section{Introduction}
One may think that since viscous friction is responsible for flow
irreversibility, the latter must disappear in the inviscid limit. On
the contrary, there exists a dimensionless measure of irreversibility
that actually grows unbounded as the viscosity goes to zero, as was
recently found for incompressible turbulence \cite{xu-pumir-etal:2014}
and as we show here for a compressible one. The reason is that when
the magnitude and scale of the flow excitation is fixed while the
viscosity is getting smaller the fluid is driven away from
equilibrium. This is a consequence of the persistence of energy
dissipation at smaller and smaller scales as viscosity tends to zero
(i.e $\Re \to \infty$). At equilibrium, time reversibility of the
statistics is manifest through detailed balance: It is equally
probable for energy to transfer between two scales of the flow in
either direction \cite{onsager:1931}. On the contrary, in a steady
state of a turbulent flow, the separation between the scale at which
energy is introduced and that at which it is dissipated is of the
order of the Reynolds number and an energy flux is formed between the
two \cite{frisch:1995}.

In other words, at equilibrium detailed balance means that excitation
and dissipation are balanced at every scale and every timescale. In
turbulence, by increasing the ratio of excitation and dissipation
scales we naturally drive the system further from equilibrium. In
light of this discussion it is clear that a measure of
time-irreversibility should be at the same time a measure of the
deviation from equilibrium. The question now is how to recover such a
measure not by looking at the {\it spatial} structure of forcing and
dissipation in the whole system (as done e.g. in
\cite{bauer-bernard:1999, eyink-drivas:2014}), but by studying the
{\it temporal} evolution of the smallest part of the flow, a single
fluid element. It requires some work to devise a measure of
time-irreversibility that can be measured using single particle
statistics: the velocity statistics are stationary implying that
velocity structure functions are invariant under $t\rightarrow -t$
\cite{falkovich-etal:2012}. Xu et al. \cite{xu-pumir-etal:2014}
suggested to study the statistics of the energy evolution of a fluid
particle, $W(t) = \frac12(u(t)^2 - u(0)^2)$, and showed that here
irreversibility is embodied as follows: a particle gains energy slowly
and loses it fast, a process they termed ``flight-crash'' events. A
measure of time-irreversibility, $\textit{Ir}$, was then constructed
by looking at the short time limit of $W(t)$, the power $p= \bf{a}
\cdot \bf{v}$ where $a$ is the Lagrangian acceleration, and it was
found that $\textit{Ir}\equiv -\left< p^3\right> /\epsilon^3 \propto
\Re^2$, where $\epsilon$ is the dissipation rate of kinetic energy.
This scaling and the skewness of the statistics of single particle
energy changes was hypothesized to originate from the ``flight-crash''
events.

In this paper, we want to apply similar techniques to measure
time-irreversibility in a \emph{compressible} flow in two
setups. ``Flight-crash'' events are expected to be present in
compressible turbulence from a general point of view: for strongly
compressible high-$\Re$ flows particles travel mostly unaffected until
colliding with other particles inside shocks \cite{beetz-etal:2008}, a
process during which they rapidly lose energy. The Burgers equation
driven by a large-scale force describes a set of distant shocks
\cite{bec-khanin:2007} and is therefore a good test bed for ideas
about ``flight-crash'' events as a source of irreversibility. It can
also serve as a simple model to explore irreversibility in strongly
compressible turbulent flows.

First we sample markers initially homogeneously distributed in the
flow. For the energy increments we find similarly skewed statistics to
those found in \cite{xu-pumir-etal:2014}, $\langle W^3 \rangle < 0$
and the scaling $\langle W^n \rangle \propto t$ which can be explained
by dominance of ``flight-crash'' events as well as, in more detail, by
modeling the fall of particles into shocks. We furthermore analyze the
power $p$ of particles and its moments, which also depend on $\Re$ in
this case.  Its scaling with $\Re$ agrees with estimates of the
relation between the shock width and viscosity as well as the
prediction from a ``flight-crash'' model. Note that this type of
sampling results in non-stationary Lagrangian velocity statistics due
to compressibility, making time-irreversibility more evident. In
particular the connection between $W(t)$ and $W(-t)$ is not simply a
sign flip.

In the second setting, we examine the long time limit, in which all
particles are located inside shocks. In this limit the velocity
statistics are stationary and time-irreversibility is less
transparent. We show that again moments of $W(t)$ and of the power can
be used to measure irreversibility and display the same qualitative
features as their short time counterparts. These results can also be
interpreted to arise from a ``flight-crash'' model where the crashes
leading to a sudden energy change are shock collisions.

The paper is organized as follows: In Sec. \ref{one} we introduce a
setting and establish a qualitative model of particle trajectories in
a turbulent flow, the so called ``flight-crash''-model. We show that
this model has a direct interpretation in compressible turbulence, as
particles crashing into shocks rapidly lose energy. To estimate
particle energy increments we invoke a steady state shock model in
Sec. \ref{two}. The predictions from this model are then compared to
numerical simulations of 1d compressible turbulence, both for energy
increments, in Sec. \ref{two} and \ref{three}, and moments of power in
Sec. \ref{four}. We then investigate Lagrangian velocity increments,
in Sec. \ref{five}, comparing results from numerical simulations to
predictions based on the competition between forcing induced
propagation and events related to shocks.  In the last part,
Sec. \ref{six}, we try to eliminate the effect of compressibility
which induces non stationary statistics of velocities by looking at
the long time statistics for a single particle. Our main result is a
numerical verification of these estimates in both setups.

\section{``Flight-Crash'' events in compressible turbulence}
\label{one}
In the following we consider the Burgers equation \cite{burgers:1974},
\begin{equation}
  \label{eq:burgers}
  v_t + vv_x - \nu v_{xx} = f\,,
\end{equation}
as a simple model for a compressible flow in 1d, where $f$ is the
forcing term. Particles at position $x(t)$ are advected with the
velocity $v(x(t),t)$, obeying the equation
\begin{equation}
  \label{eq:particle}
  \frac{\mathrm{d}x(t)}{\mathrm{d}t} = v(x(t),t),\qquad x(0)=x_0\,.
\end{equation}
We furthermore denote the particle velocity with $v(x(t),t)=u(t)$.
The quantities that are of interest to us are the moments of kinetic
energy differences along trajectories of fluid elements, $\langle
W(t)^n\rangle $, as well as moments of the instantaneous power for
fluid elements distributed homogeneously, $\left<(a(0)v(0))^n\right>$.

All numerical simulations carried out for this work integrate equation
(\ref{eq:burgers}) in time, using a second order stochastic
Runge-Kutta algorithm \cite{honeycutt:1992} in time and fast Fourier
transforms for all space derivatives. The tracer particles are
  integrated with the same time-marching algorithm and a second-order
  field interpolation. For stochastic forcing we employ both
Brownian noise with $\delta(t)$ correlation in time and finitely
correlated noise with correlation time $T_f$ implemented as an
Ornstein-Uhlenbeck process in Fourier space \cite{eswaran-pope:1988},
depending on the physical requirements. The Reynolds number is varied
in the range $10^1<\Re<10^4$. Different $\Re$ are achieved by
modifying $\nu$ while retaining the shape of the forcing. The
implementation uses graphics processing units (GPU) and the CUDA
framework \cite{cuda:2014} for speedup and allows us to reach
$10^8$--$10^{10}$ computational steps, which amounts to approximately
$10^4$--$10^5$ integral times per simulation.

In Burgers turbulence \cite{bec-khanin:2007}, a finite number of shock
structures, i.e. subsets with a large $\partial_x v$, emerge with a
density $\rho = 1/L$, where $L$ is the forcing correlation
length. These structures capture surrounding particles. Regions
between shocks are comparably smooth, and the relative motion between
particles in those regions and the neighboring shocks is approximately
ballistic. If particles are injected with a uniform distribution at
$t=0$, almost all of them are initially located in the smooth regions
between the shocks. Each particle then undergoes a ballistic motion
with respect to the nearest shock until crashing into it.  In terms of
the energy difference at time $t$, there are therefore two types of
events that one expects to contribute: The most common events are
those where the particle gains energy slowly due to the forcing far
away from any shock; the rare events occur when the particle enters a
shock during the time $t$, losing a large amount of energy.  The
latter events are naturally interpreted as ``flight-crash''-events and
give the largest contribution to the energy difference.

Following the derivation of this model in the incompressible case, let
us evaluate the contribution of the rare events in which energy is
lost giving $W<0$. First we use the decomposition $\langle
W(t)^n\rangle \approx \langle \left(v_{\text{rms}}(u(t) -
  u(0))\right)^n\rangle $ and apply the estimate $(u(t)-u(0))\approx
(v(r)-v(0))$ with $r=v_{\text{rms}} t$ the initial distance between
the particle and the shock it enters for sufficiently short times.
Then the Eulerian scaling $\langle (v(r)-v(0))^n\rangle \propto r$
implies the scaling $\langle W^n\rangle \propto t$ for $n>1$. As the
Eulerian moments are not self similar the energy difference is not
self similar either. Note that although we also get that $\langle
W^3\rangle\propto t $, as in the prediction in the incompressible
case, this is not a general feature of a ``flight-crash''-type of
argument but rather depends on the scaling of the third order Eulerian
structure function.

Our estimate for $W$ also allows us to obtain predictions for the
scaling of the power with the $\Re$. The scaling we derived above is
expected to hold for times $t\geq \tau_{\eta}$, $\tau_{\eta}$ being
the viscous time scale after which the internal structure of the shock
does not matter. At times $t \leq \tau_{\eta}$ a Taylor expansion in
time implies $\left<W^n\right> \propto \left<p^n\right> t^n$. At
$t=\tau_{\eta}$ these two scalings should match, giving
$\left<p^n\right> \propto \tau_{\eta}^{1-n}\propto \Re^{n-1}$. Note in
passing that there is also an upper bound for our prediction for
$W(t)$, $t<T_L$, with $T_L$ the typical time related to the forcing
scale $L$. Up to this time the spatial variation of the forcing is not
yet felt by the particle.  For forcing with a finite time correlation
there is an additional time scale, which we take to be of the order of
$T_L$.

To summarize, according to the ``flight-crash'' model we expect to find
$\left<p^n\right> \propto \Re^{n-1}$ for the power moments and
$\left<W^n\right> \propto t$ for the energy difference. We will
substantiate these estimates in the following sections.

\section{Moments of energy differences along fluid trajectories}
\label{two}
\begin{figure}[tb]
 \begin{center}
    \includegraphics[width=200pt]{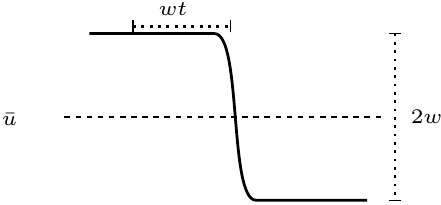}
  \end{center}
  \caption{Prototypical shock solution of the Burgers equation. The
    shock height is $2w$ and the shock velocity is $\bar u$. Particles
    on the brink of the shock enter the shock and lose energy. The
    amount of particles that enter the shock in time $t$ from the left
    or right side is $wt$.}
  \label{fig:shock}
\end{figure}

Let us study the energy differences in more detail. From $u^2 = u_0^2
+ 2 \int uf \,dt + 2 \nu \int u u_{xx} \,dt$ we can write $W(t) = \int
uf \,dt + \nu \int u u_{xx} \,dt = F(t) + D(t)$ with $D(t) = \nu \int
u u_{xx} \,dt$ being the amount of energy dissipated at the particle position
and $F(t)$ the contribution to the energy from the forcing. The
forcing term can be estimated as follows: initially, due to
  the balance between forcing and dissipation $\left<uf\right> =
  \epsilon$, meaning that its average is $\left<F\right>=\int
\left<uf\right> \,dt = \epsilon t+ O(t^2)$ and for the higher moments
we can use $\left<F^n\right> = O(t^n)$.  To evaluate the
  dissipation term we cannot use the latter argument, relying on a
  Taylor expansion, for times $t>\tau_{\eta}$
  \cite{frishman-falkovich:2014}. Instead, anticipating that the main
contribution to $D$ comes from shocks, we will use the simplest model
of a shock to get estimates on $D$. Within this model we will compute
the energy loss along particle trajectories for a given shock and then
average over the shock parameters \cite{frishman-falkovich:2014}.

\begin{figure}[tb]
  \begin{center}
    \includegraphics[width=\columnwidth]{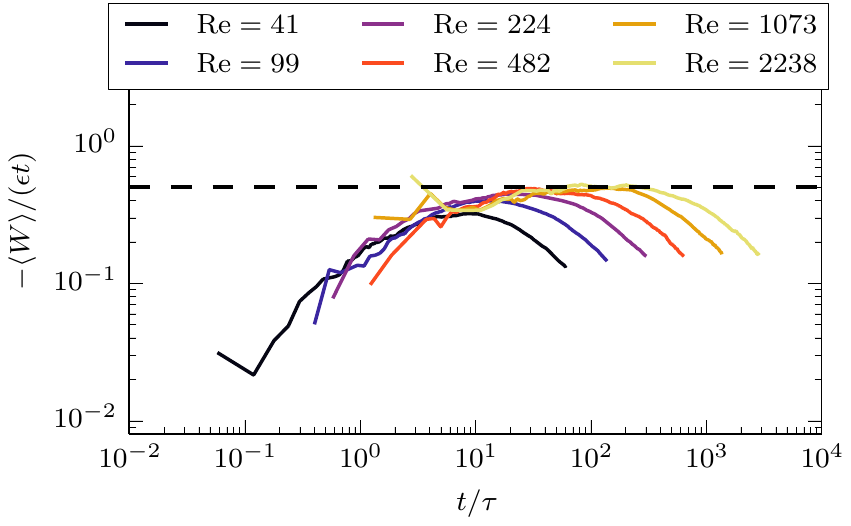}
  \end{center}
  \caption{(Color online) $-\langle  W(t) \rangle $ versus $t/\tau$, compensated by
    $\epsilon$ and $t$, for stationary Burgers turbulence, considering
    particles that start at a random position at $t=0$. Particles on
    average lose energy linearly in time, in accordance to the
    analytical estimate in equation~(\ref{eq:W}).}
  \label{fig:W}
\end{figure}

Consider a prototypical shock solution of the Burgers equation, as
depicted in figure~\ref{fig:shock}. Let the shock height be $2w$ and
the shock velocity be $\bar u$.  Then, particles entering from left
and right lose different amounts of energy:
\begin{align*}
  D_1 &= -\frac12 w^2 - w\bar u\\
  D_2 &= -\frac12 w^2 + w\bar u.
\end{align*}
The probability of a particle to enter a shock during time t,
  from either side of it, is $wt\rho$ where the shock density is
  related to the forcing scale $\rho=1/L$.  Thus $\langle D \rangle =
-\langle w^2 \frac{w t}L \rangle = \frac{3}2 \epsilon t$
\cite{falkovich2011fluid}. Finally, adding the contribution from the
forcing one obtains
\begin{equation}
  \label{eq:W}
  \left< W \right> = -\frac12 \epsilon t.
\end{equation}
for $\tau_{\eta}<t<T_L$. This estimated scaling in $\epsilon$ and $t$,
as well as the pre-factor, agree well with numerical simulations using
a white in time correlated forcing, as shown in figure~\ref{fig:W}.
Corroborating this result for a finite correlated forcing requires a
significant increase of the statistics as well as of the correlation
time $T_f\approx T_L$ compared to those we used. While we did not
attempt to do so, the partial results we obtained did not seem to
contradict (\ref{eq:W}).  Equation (\ref{eq:W}) demonstrates the main
difference between Burgers and incompressible turbulence --- the
Lagrangian energy is not stationary; for short times, fluid elements
lose energy linearly in time, on average, as opposed to $\langle W(t)
\rangle =0$ in the incompressible case.

\begin{figure}[tb]
  \begin{center}
    \includegraphics[width=\columnwidth]{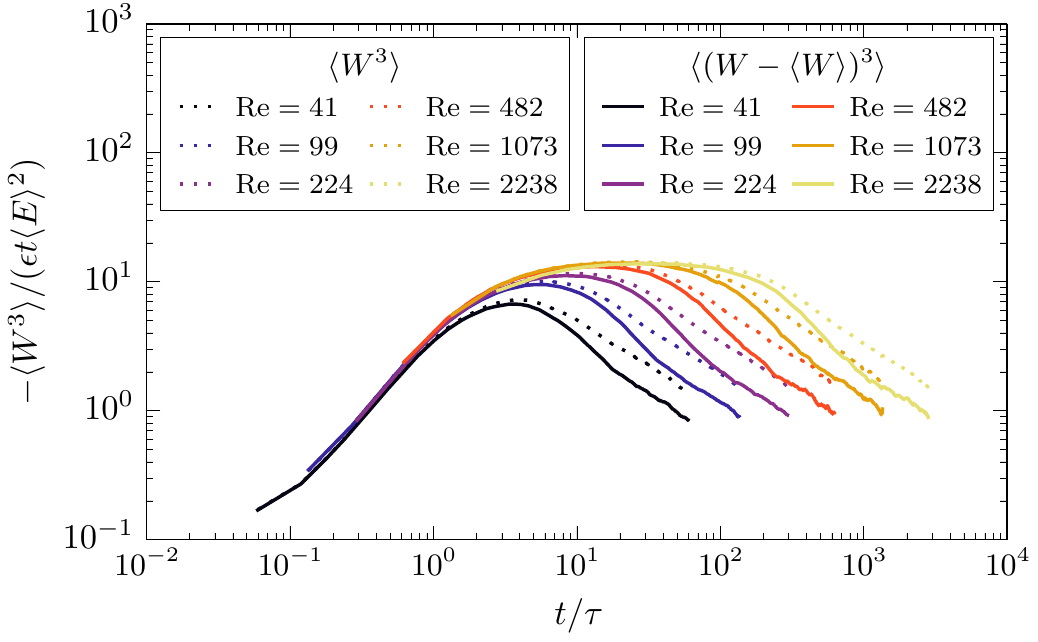}
  \end{center}
  \caption{(Color online) In the inertial range, the centered moment
    (solid) $\langle (W-\langle W \rangle )^3 \rangle $ scales similar
    to to $\langle W^3 \rangle $ (dotted), which is in accordance with
    the analytical estimate in equation~(\ref{eq:W3}).}
  \label{fig:W3_cumulant}
\end{figure}

Turning to $\langle W^3 \rangle$, it is dominated by $\langle D^3
\rangle $ in the inertial range since the terms involving the forcing
are sub-dominant by at least one factor of $t/T_L$. Thus one obtains
\begin{equation*}
  \left< W^3 \right> \approx \left< D^3 \right> = \left<\frac{wt}{L} \left( D_1^3 + D_2^3 \right)\right> = -\frac{t}{4L}\left<12 \bar u^2 w^5 + w^7\right>.
\end{equation*}
For white in time forcing this equation can be re-expressed in terms
of Eulerian velocity moments using
\begin{equation*}
\left<v^{4}\right>=\frac{70 \left<\bar{u}^4 w^3\right>+84 \left<\bar{u}^2 w^5\right>+6 \left<w^7\right>}{105 L \epsilon }
\end{equation*}
from \cite{e-vandeneijnden:2000} to substitute for $\left<
  w^7\right>$ and thus obtain the bound
\begin{equation}
  \label{eq:W3}
  -\left< W^3 \right> < \frac{35}2 \epsilon t \frac{\left< v^4 \right>}4\,.
\end{equation}

We present $\left< W^3 \right>$ for white in time correlated forcing
in figure~\ref{fig:W3_cumulant} (dotted). The expected scaling with
$t$ and $\epsilon$ is supported by the numerical results, the plateau
increasing in length for growing Reynolds numbers. Furthermore it is
evident that the bound in Eq.~(\ref{eq:W3}) is satisfied. We reproduce
the same qualitative behavior with finite-time correlated
forcing. This result is much easier to obtain than the one for
$\left<W\right>$ as the forcing enters only sub-dominant terms in
$\left<W^3\right>$ in the inertial interval.

Similarly, for a general moment $\langle W^n \rangle $ we expect
\begin{equation*}
  \left< W^n \right> \approx t
\end{equation*}
regardless of $n$, the dissipative term giving the dominating
contribution for all $n>1$, $\langle W^n \rangle \approx \langle D^n
\rangle$.

\section{Centered moments of energy differences}
\label{three}

For a compressible flow there is an obvious source of
irreversibility for an initially homogeneous distribution of particles
in space, since the particle density changes in time. This is why,
unlike for the incompressible flow, already the first moment of $W$ shows
irreversibility: $\langle W\rangle <0$.
One might wonder whether by subtracting this average, i.e looking at
centered moments $\langle (W-\langle W\rangle )^n\rangle $, it is
possible to eliminate the footprints of irreversibility. In other
words, subtracting the mean brings the situation closer to the
incompressible case, with a random variable whose mean vanishes and
its third centered moment being non-zero demonstrates
irreversibility. This is indeed the case but the irreversibility in
time is still dominated by the (linearly in time) increasing
probability for a particle to crash into a shock.

In fact, we expect that the leading contribution would come from
$\langle W^3\rangle \propto t$ rather than from $\langle W\rangle ^3
\approx O((t/T_L)^2)\langle W^3\rangle $ and $\langle W^2\rangle
\langle W\rangle \approx O((t/T_L))\langle W^3\rangle $.

We can also use the above model to estimate the difference between
$\langle (W-\langle W\rangle )^3\rangle $ and $\langle W^3\rangle $:
\begin{equation}
  \left<(W-\left<W\right>)^3\right> =
  \left<W^3\right>-3\left<W^2\right>\left<W\right>+2\left<W\right>^3
\end{equation}
Then, subtracting this from $\langle W^3\rangle $ to leading order in
$t/T_L$ we have
\begin{equation}
 - \left<W^3\right>+\left<(W-\left<W\right>)^3\right>\approx\frac{3}{2}
  \left<D^2\right>\epsilon t
\end{equation}
after using $-3\left<W^2\right>\left<W\right> \approx
-3\left<D^2\right>\left<W\right>=3/2 \left<D^2\right>\epsilon t$.  We
therefore expect for times $t \approx T_L$, when this sub-leading term
becomes visible, that the negative of the centered moment would lie
lower than $-\left< W^3\right>$ . As shown in
figure~\ref{fig:W3_cumulant} (solid lines), this is consistent with
what is observed in the numerical simulation.

\section{Moments of power}
\label{four}
\begin{figure}
  \includegraphics[width=0.49\columnwidth]{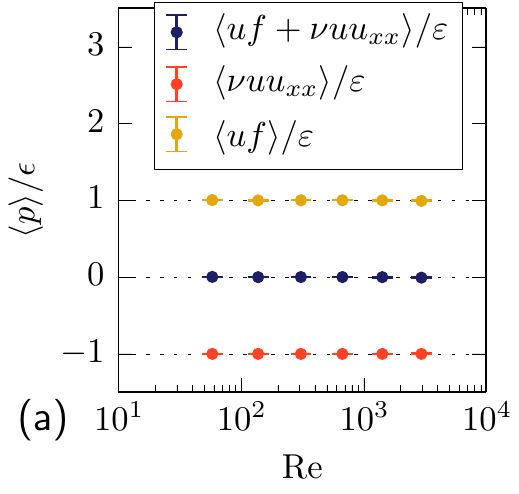}
  \includegraphics[width=0.49\columnwidth]{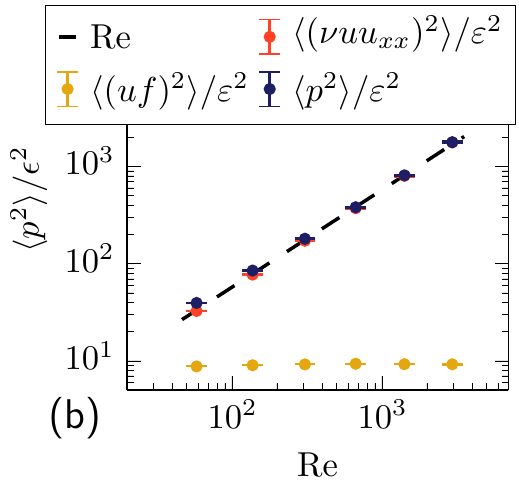}
  \caption{(Color online) (a) The first moment of power, $\langle p
    \rangle $, is equal to zero. Note how forcing and dissipation
    exactly cancel each other. (b) The second moment of power,
    $\langle p^2 \rangle $, scales like $\Re$. The dissipative term
    dominates the forcing term.\label{fig:p}}
\end{figure}

\begin{figure}
  \includegraphics[width=\columnwidth]{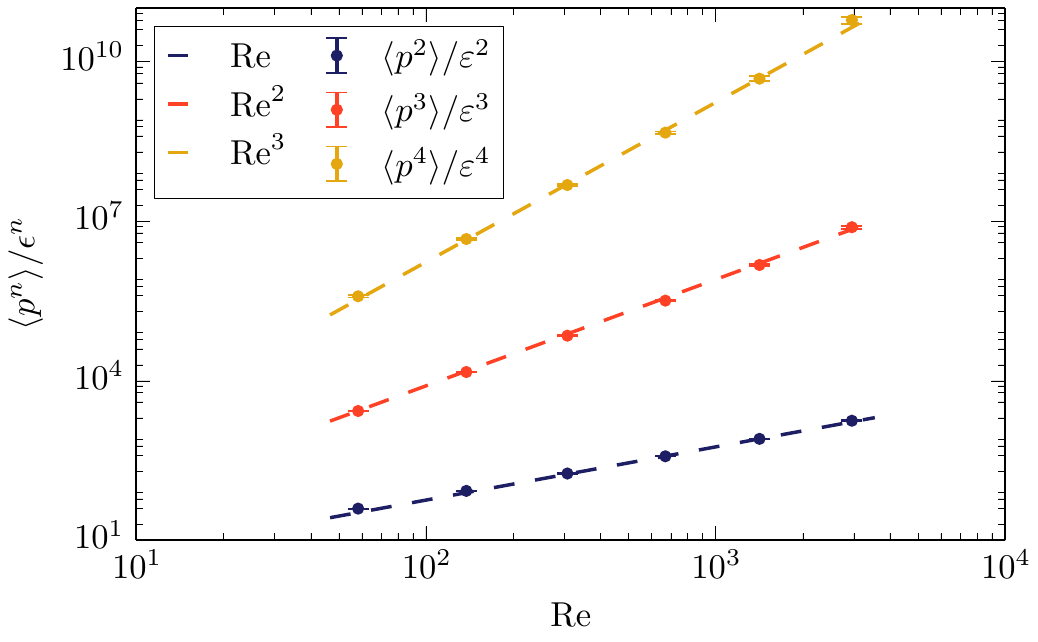}
  \caption{(Color online) Higher moments of power, $\langle p^n
    \rangle $, scale like $\Re^{n-1}$.\label{fig:p_higher}}
\end{figure}
In this section we explore a measure of irreversibility that is a
direct consequence of the existence of an energy cascade.  We consider
$\langle p^3\rangle =\langle \left(\frac{dE}{dt}\right)^3\rangle $
which for a time-reversible system would be equal to zero. Note that
the first moment, $\langle p \rangle=0$, due to the balance between
dissipation and forcing.  For white in time forcing such quantities
are ill defined as they correspond to time derivatives at $t=0$. We
therefore use forcing with a finite correlation time in the numerical
simulations presented here.

To obtain a prediction for the scaling of $\langle p^n\rangle $ with
the $\Re$ we will use a dimensional reasoning of sorts. First, in
general, we can use the Burgers equation to write
\begin{equation}
  \label{eq:pn_}
  \left< p^n\right>=\left< \left(v \frac{dv}{dt}\right)^n\right>=\left<\left[ v\left(f+\nu v_{xx}\right)\right]^n\right>.
\end{equation}
Now, any average including $\nu v v_{xx}$ is concentrated on the shock
locations (or places with very large gradients for finite
viscosity). These are small regions of thickness of $\nu/v_{rms}$
where the velocity spatial gradient is proportional to $
v_{rms}^2/\nu$. Thus, for $n\geq 2$ terms including the forcing are
sub-dominant to $\langle \left( \nu v v_{xx}\right)^n\rangle $ by at
least one factor of Reynolds number, $\Re=v_{rms} L/\nu \propto
\epsilon^{1/3} L^{4/3}/\nu$. In particular
\begin{equation}
  \label{eq:pn}
  \left< p^n \right> \approx \left< \left( \nu v
  v_{xx}\right)^n\right> \propto \frac{1}{L}\nu^n
  \left(\frac{v_{rms}^4}{\nu^2}\right)^n\frac{\nu}{v_{rms}} \propto \epsilon ^n
  \Re^{n-1}.
\end{equation}
These dimensional estimates, coinciding with the predictions of the
``flight-crash'' model, are supported by numerical experiments. For
the first moment, as shown in figure~\ref{fig:p} (left), the forcing
and dissipation terms cancel each other. Figure~\ref{fig:p} (right)
depicts the scaling of the second moment of power, $\langle
p^2\rangle$ proportional to $\Re$. Note also that the dissipative
term, $\langle \left( \nu v v_{xx}\right)^2\rangle $ dominates the
forcing term $\langle \left( v f\right)^2\rangle $, the latter being
unaffected by changes in $\Re$. This is a very different situation
from that in 2d and 3d incompressible turbulence, where the leading
$\Re$ dependence of the power comes from pressure terms. As shown in
figure \ref{fig:p_higher}, the second to fourth moment of power scale
like $\Re^{n-1}$, in accordance with the dimensional estimate of
(\ref{eq:pn}).

\section{Lagrangian velocity increments}
\label{five}

\begin{figure}[tb]
  \begin{center}
    \includegraphics[width=\columnwidth]{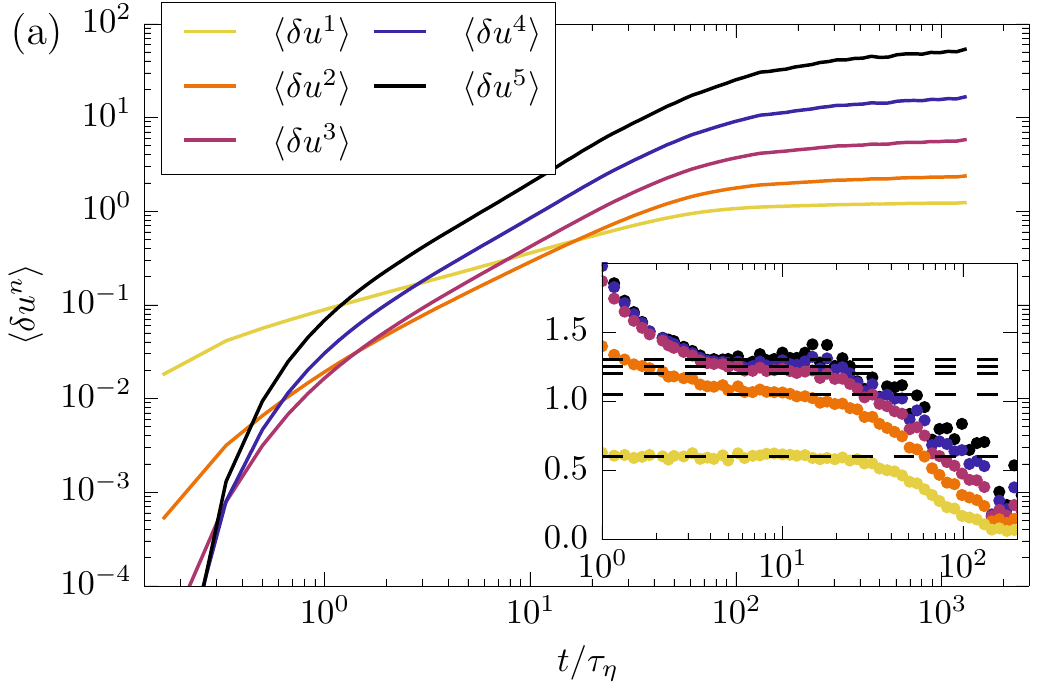}\\
    \includegraphics[width=\columnwidth]{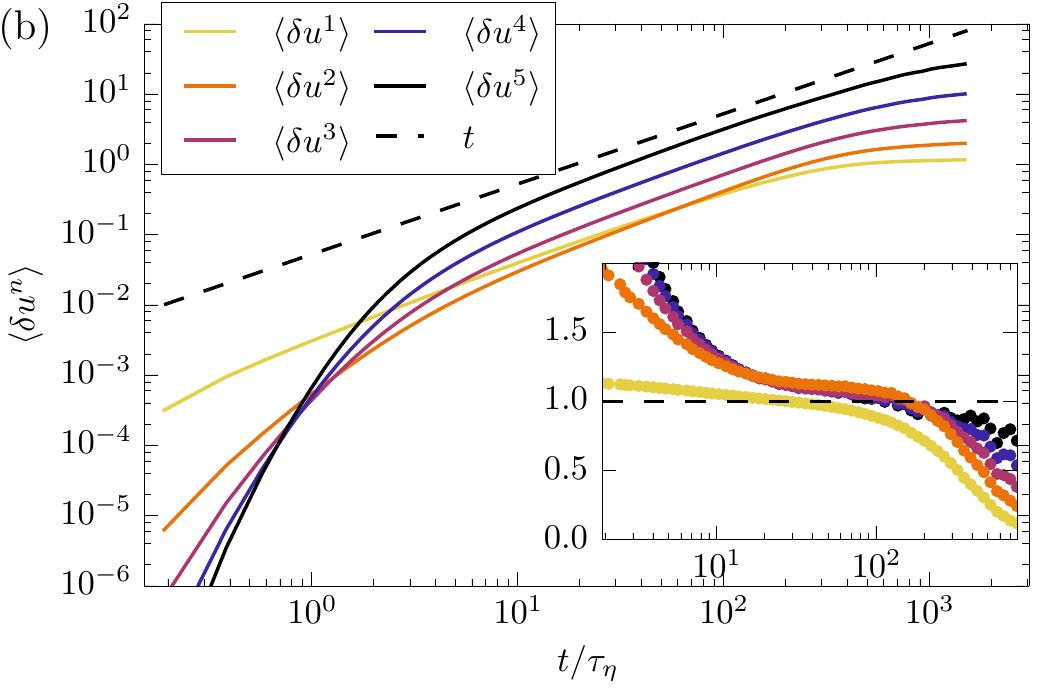}
  \end{center}
  \caption{(Color online) Lagrangian velocity increments for white in
    time forcing (a) and finite forcing correlation time (b),
    the corresponding local slopes are presented in the respective
    insets. For the white in time forcing the exponents from
    Table~\ref{tbl:deltav} are marked by dashed lines}
  \label{fig:increments}
\end{figure}

As the velocity statistics are non-stationary one may expect to be
able to determine the direction of time from the sign of odd moments
of velocity differences. In fact, this is not possible since
  such odd moments are zero: using the invariance of the system under
  space reflection $\left< (u(t;x_0)-u(0;x_0))^n\right>=(-1)^n\left<
  (u(t;-x_0)-u(0;-x_0))^n\right>$, and the independence of the average
  on $x_0$, the initial particles position, completes the proof.  Of
course, non-stationary statistics imply that already the even moments
behave differently for positive and negative times ($t=0$
corresponding to a homogeneous particle distribution). In particular
backward in times velocity differences are determined solely by the
forcing, as particles do not encounter shocks. On the other hand, as
we will show, shocks provide a significant contribution forward in
time.

The dependence on time of
$\left<\left|u(t)-u(0)\right|^n\right>=\left<\delta u^n\right>$ can be
deduced similarly to that of Eulerian velocity differences in this
system. There are two competing time scalings which imply
bi-fractality. Events where particles do not fall into shocks, which
occur with probability $O(1)$, change their velocity diffusively or
ballistically depending on whether the forcing is short or finite
correlated in time. This gives $\delta u^n \propto t^n$ for finitely
correlated and $\delta u^n \propto t^{n/2}$ for delta correlated
forcing. On the other hand there are the events where particles fall
into shocks, the probability for which scales linearly with time and
where the velocity difference is $\delta u^n \propto O(1)$.  This
implies that for times $\tau_{\eta} \ll t \ll T_L$, for white in time forcing we expect

\[\left<\delta u^n\right>\propto  \left\{
  \begin{array}{ll}
    t^{n/2} & , n<2\\
    t & , n \geq 2
  \end{array}
\right.
\]
while for forcing with a finite correlation time
\[\left<\delta u^n\right>\propto  \left\{
  \begin{array}{ll}
    t^{n} & , n<1\\
    t & , n \geq 1.
  \end{array}
\right.
\]

Figure~\ref{fig:increments} shows the results from numerical
simulations, for the forcing correlated both short and long in time,
largely agreeing with the prediction above.  The measured local slopes
$d\ln \langle\delta u^n\rangle/d\ln t$ are presented in the inset. As
a guidance for the eye, we have marked the approximate scaling
exponents in the inset of Figure~\ref{fig:increments} for the white in
time forcing.  Their values are summarised in Table
\ref{tbl:deltav}. For the long correlated forcing, although the local
slopes are close to $1$, there are no clear plateaus, possibly due to a
longer influence of the dynamics at $t \approx \tau_{\eta}$ on the
inertial range.

The deviations of the measured local slopes from our prediction
apparent in Table \ref{tbl:deltav} and Figure~\ref{fig:increments} are
probably a finite $Re$ effect as well -- in the inertial range both
competing time scalings are present for all $n$, a single scaling
becoming dominant only in the limit $Re\to \infty$. Indeed, the best
agreement is observed for $n$ where the two terms are of the same
order: $n=1$ for long correlated forcing and $n=2$ for white in time
forcing.  This would also explain why the white in time local slopes
are further from the prediction than the long correlated ones, the two
competing terms being closer to each other for the former.

It is worth noting that $\zeta_2=1$ is also the prediction for
incompressible turbulence in 2d and 3d obtained by dimensional
arguments or the multi-fractal phenomenology
\cite{Biferale2004,BoffetaMazzino2008}. Such a relation, however, was
never clearly observed either numerically or experimentally
\cite{falkovich-etal:2012,BoffetaMazzino2008}. For the Burgers
equation it can also be derived on dimensional grounds, as well as by
using the Lagrangian multi-fractal phenomenology. The latter relates
the Eulerian scaling to the Lagrangian one by assuming that the time
elapsed can be related to the distance travelled via $t \propto
r/\delta_{r} u$ and that $\delta u(t) \propto \delta_r u $ where
$\delta_r u=v(r,t)-v(0,t)$ is the Eulerian velocity difference. Then
it is predicted that $\zeta_n=\min\limits_h
\left[\frac{nh-D(h)+d}{1-h}\right]$ with $D(h)$ the Eulerian fractal
dimension.

For the Burgers equation on shocks $h=0$, $D(0)=0$ and everywhere else
$h=1$, $D(1)=1$, which gives the correct prediction for $n\geq
2$. Indeed, the above assumptions are satisfied for the Burgers
equation for $n\geq 2$ as shock events control the statistics: due to
the presence of the shock $\delta u(t)\propto \delta_r u \propto O(1)$
and since the particle moves ballistically relative to the shock $t
\propto r$. For $n<2$ while $\delta u(t)\propto \delta_r u$ the
Eulerian velocity difference tells nothing about the distance
travelled $r$, as demonstrated by the dependence of $\zeta_1$ on the
temporal correlation of the forcing.

For incompressible turbulence, while the Lagrangian multi-fractal
phenomenology leads to a good fit in 3d
\cite{Arneodo2008,toschi2009,bec-bitane2013}, the assumption $t
\propto r/\delta_{r} u$ cannot be universally exact
\cite{KampsFriedrich2009}. In particular, thinking of averages as a
weighted sum over events, different events may dominate the average
depending on the quantity one considers, and while the relation $t
\propto r/\delta_{r}u$ may work well for some events it can fail for
others.  Indeed, to obtain the scaling of the energy difference,
dominated by flight crash events, $t \propto r$ and $\delta u(t)
\propto \delta_r u $ were used in \cite{xu-pumir-etal:2014}.  An
elegant way to amplify these same events was recently introduced in
\cite{LevequeNaso2014} where new longitudinal Lagrangian velocity
increments were defined and measured instead of energy differences,
revealing that the projection on the direction is the main
ingredient. Then, assuming $t \propto r$ and $\delta u(t)_L \propto
\delta_r u $ implies that the Eulerian and longitudinal Lagrangian
velocity moments should have the same scaling exponents. This was
verified for the third and second order velocity moments in
\cite{LevequeNaso2014}. It is however unclear why the assumption $t
\propto r$ should hold.  In this context our observations for the
Burgers turbulence may provide some insight: if the change in the
particles velocity is due to transition at time $t$ into a region with
a different velocity scaling then the distance travelled should be
determined by the relative velocity between the two regions, i.e $t
\propto r/\delta_{r} u$. On the other hand, the distance travelled by
a particle within a region with a single scaling is detached from
$\delta_{r} u$.

\begin{table}[tb]
  \centering
  \begin{tabular}{l|rrrrr}
    $p$ & 1 & 2 & 3 & 4 & 5\\
    \hline
    $\zeta_p^{\text{white}}$ & 0.6 & 1.05 & 1.2 & 1.25 & 1.3
  \end{tabular}
  \caption{Scaling exponents of Lagrangian
    velocity increments for white in time forcing.}
  \label{tbl:deltav}
\end{table}

\section{Long time statistics}
\label{six}

\begin{figure}[tbp]
  \includegraphics[width=\columnwidth]{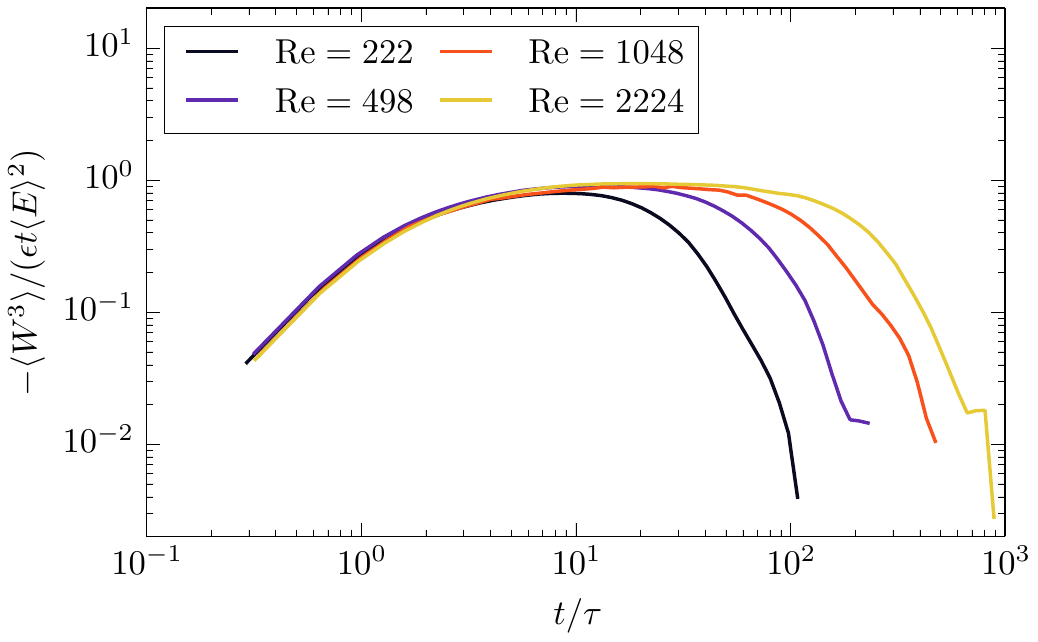}
  \caption{(Color online) Single particle statistics for long
    times. Again, $-\langle W^3 \rangle \propto \epsilon
    t$.\label{fig:w3_single}}
\end{figure}
\begin{figure}[tbp]
  \includegraphics[width=\columnwidth]{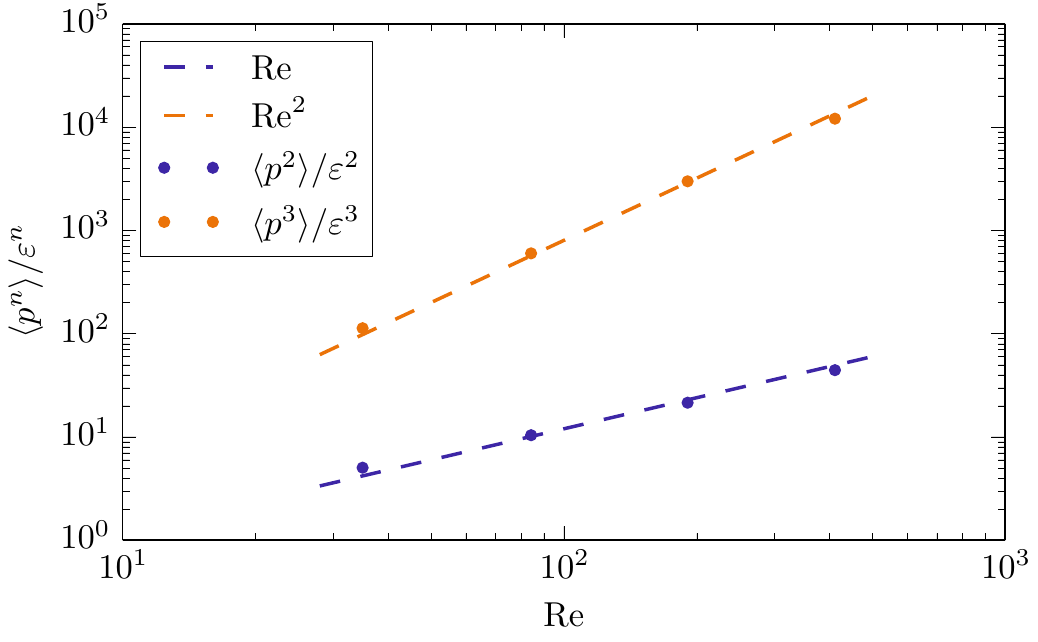}
  \caption{(Color online) The moments of power, $\langle p^n\rangle$,
    for single particle statistics for long times. We recover the
    scaling $\langle p^n\rangle \sim
    \textrm{Re}^{n-1}$. \label{fig:p_noreset}}
\end{figure}
As long as the Lagrangian velocity statistics, starting from a
homogeneous distribution of particles initially, do not reach a steady
state, the irreversibility of the system cannot be attributed solely
to the existence of an energy cascade. It is this somewhat trivial
component of the irreversible dynamics which we wish to eliminate when
considering long time statistics.  For long times, all particles
accumulate inside shocks, and since any two shocks eventually
  merge, for very long times, all particles reach the same position
in the same shock. Therefore, observing long time statistics is
equivalent to considering only a single particle in the entire flow.
As new shocks are created with time and, in the spirit of our
  considerations in the previous sections, we expect the main
  contribution to come from shock collisions, particle statistics in
this regime can also be seen as shock-interaction statistics. We we
will perform similar measurements for particles as above, this time
arbitrarily defining some $t=0$ when stationary particle statistics
are reached, instead of starting with a homogeneous particle
density. In contrast to the short-time case it is much harder to
obtain good statistics, since the flow is only sampled at a single
particle position. We therefore resort to the more robust method of
estimating the moments of power $\langle p^n \rangle$ by finding the
plateau of $\langle W^n(t) \rangle/t^n$ for short times, instead of
evaluating $\langle dE/dt \rangle$ directly. We furthermore restrict
the range of $\Re$ to lower values for the power statistics, as
obtaining converging results for higher $\Re$ becomes
prohibitive. Reaching stationary statistics implies in particular
$\langle W \rangle =0$, which we indeed observe.  We obtain that
$\langle W^3 \rangle <0$, depicted in figure \ref{fig:w3_single}. It
turns out that the scaling is very similar to that at short times,
with $\langle W^n \rangle \propto t$ and $\langle p^n \rangle \propto
\Re^{n-1}$. The corresponding results from numerical simulations are shown in
figure~\ref{fig:p_noreset} for the second and third moment of power.

We believe that a qualitative explanation for this behavior can be
given in terms of shock collisions. Shock collisions are rare events,
where the velocity of a given shock is changed by an order one
factor. Between such events the shock slowly changes its velocity due
to the forcing. In this sense we recover again a ``flight-crash''
picture: The linear scaling with $t$ of $\langle W^n \rangle $ is due
to the probability for a shock collision. It is proportional to the
probability to encounter a shock during time $t$, which scales like
$t$. The scaling of the power moments can then again be derived by
matching the scaling of $\left<W^n\right>$ for times $t \leq
\tau_{\eta}$ and $t \geq \tau_{\eta}$ at $t=\tau_{\eta}$. We note that
not every shock collision results in an energy loss. Energy must be
lost on average though in order to balance the energy gain from
forcing and obtain the stationary state $\left<W\right>=0$.

\section{Conclusion}
We have studied time-irreversibility as deduced from the statistics of
a single element, a fluid marker, in a compressible turbulent flow.
Transferring the ideas of Xu et al. \cite{xu-pumir-etal:2014} to
Burgers turbulence, we measured Lagrangian energy differences and
instantaneous power statistics, and demonstrated the ability of the
``flight-crash'' model, suggested therein, to explain our results. 
From the point of view of particles, compressibility itself introduces an additional element of time-irreversibility in the form of shock structures. Therefore,  we consider two different regimes: First we
consider the trajectory of a particle starting at a random
position. Here, we estimate the scaling $\langle W(t)^n \rangle
\propto t$ and $\langle p^n \rangle \propto \textrm{Re}^{n-1}$ by
invoking a steady-state shock model. Our numerical simulations confirm
these predictions for the form and the pre-factor of $\langle W(t)
\rangle$ as well as the general scaling of $\langle W(t)^n \rangle$
and $\langle p^n \rangle$. Secondly, we examine long-time statistics,
where all particles have accumulated in shocks. This regime can be
interpreted as shock-interaction statistics. The ``flight-crash''
picture then applies to the motion of shocks themselves, as they gain
energy slowly until hitting another shock, leading to a rapid loss of
energy on average. These considerations are again backed by numerical
simulations, consistent with $\langle W^n(t) \rangle \propto t$ and
$\langle p^n \rangle \propto \textrm{Re}^{n-1}$.

\section*{Acknowledgments} We thank Alain Pumir for suggesting also
using long time correlated forcing. The work of T.G. was partially
supported through the grants ISF-7101800401 and Minerva Coop
7114170101. A.F. is supported by the Adams Fellowship Program of the
Israel Academy of Sciences and Humanities. G.F. is supported by the
BSF and the Minerva Foundation with funding from the German Ministry
for Education and Research.

T.G and A.F contributed equally to this work.

\end{document}